\documentclass[11pt]{llncs}
\usepackage[letterpaper]{geometry}

\usepackage{amsmath,latexsym,xspace}
\ifx\pdfcompresslevel\undefined
\usepackage[dvips]{graphicx}
\else
\usepackage[pdftex]{graphicx}
\fi

\newcommand{\xx}[2]{\ensuremath{\def\arraystretch{0.5}\left(\begin{array}{r}%
\scriptscriptstyle #1\\\scriptscriptstyle #2\end{array}\right)}}
\newcommand{\xxx}[3]{\ensuremath{\def\arraystretch{0.5}\left(\begin{array}{r}%
\scriptscriptstyle #1\\\scriptscriptstyle #2\\\scriptscriptstyle #3\end{array}%
\right)}}

\newtheorem{Definition}{Definition}
\newtheorem{Proposition}{Proposition}
\newtheorem{Claim}[Proposition]{Claim}
\newtheorem{Corollary}[Proposition]{Corollary}
\newcommand{\valeur}[3]{\ensuremath{\mathcal{V}\xxx{#1}{#2}{#3}}}
\newcommand{\W}[4][W]{\ensuremath{\mathcal{#1}_{#2,#3}^{(#4)}}}
\newcommand{\WW}[2]{\ensuremath{\mathcal{W}_{#1,#2}}}
\newcommand{\ene}{\mathbf{N}}
\newcommand{\zed}{\mathbf{Z}}

\newcommand{\vzero}{\ensuremath{\vectk{0}}}
\newcommand{\vun}{\ensuremath{\vectk{1}}}
\newcommand{\vki}{\ensuremath{\vectk{i}}}
\newcommand{\vkx}{\ensuremath{\vectk{x}}}
\newcommand{\vixu}{\ensuremath{\vki-\vkx-\vectk{1}}}
\newcommand{\vs}{\ensuremath{t_0\cdot\vun-\vectk{u}}(t_0)}
\newcommand{\lcms}[1][\null]{\ensuremath{\lcm(1,\dots,\cardstates)^{%
      i_1+\dots+i_k#1}}}

\newcommand{\D}[2]{\ensuremath{\mathcal{D}_{#1}^{#2}}}
\newcommand{\Dg}[1]{\ensuremath{\mathcal{D}_{#1}}}
\newcommand{\cardstates}{\ensuremath{|\states|}}
\newcommand{\xinVMred}{\vectk{x}\in\Vmoore\setminus\{\vectk{1}\}}
\newcommand{\xinVM}{\vectk{x}\in\Vmoore}
\newcommand{\fff}[1]{f(\underbrace{\D{\vixu}{#1},\dots}_{\xinVM})}
\newcommand{\Vtrellis}{\ensuremath{V_{\text{trellis}}}}
\newcommand{\Vmoore}{\ensuremath{V_{\text{Moore}}}}

\newcommand{\vectk}[1]{\ensuremath{\mathbf{#1}}}
\newcommand{\cellk}[2]{\ensuremath{\left(#1,#2\right)}}
\newcommand{\statk}[2]{\ensuremath{\left\langle#1,#2\right\rangle}}
\newcommand{\states}{\ensuremath{\mathcal{S}}}

\newcommand{\Vsignal}[1][\null]{$V$\!-signal#1\xspace}

\DeclareMathOperator*{\lcm}{lcm}

\author{Jean-Christophe Dubacq\inst{1} \and V\'eronique Terrier\inst{2}}
\tocauthor{Jean-Christophe Dubacq (Universit\'{e} de Paris-Sud),
V\'eronique Terrier (Universit\'{e} de Caen)}

\institute{Universit\'{e} de Paris-Sud,
LRI, B\^{a}timent 490,\\
F-91405 Orsay Cedex, France
\and
GREYC, Campus II, Universit\'e de Caen,\\
F-14032 Caen Cedex, France}

\title{Signals for Cellular Automata in dimension 2 or higher}
\titlerunning{Signals for CA in dimension 2 or higher}

\begin{document}
\maketitle
\pagestyle{headings}
\begin{abstract}
  We investigate how increasing the dimension of the array can help to
  draw signals on cellular automata. We show the existence of a gap of
  constructible signals in any dimension. We exhibit two cellular
  automata in dimension 2 to show that increasing the dimension allows
  to reduce the number of states required for some constructions.
\end{abstract}

\section{Introduction}

Cellular automata (CA) are simple mechanisms that appear in many fields.
They are best described as simple cells regularly arranged in an array of
dimension $k$. All these cells have a finite number of states, and
change all at the same time (synchronously) of state according to the
same rules, looking at their neighbors. Physical systems containing
many discrete elements with local interactions are conveniently
modeled as cellular automata, such as dendritic crystals growth,
evolution of biological populations...

Introduced by von Neumann in~\cite{vN} to study self-reproduction,
cellular automata emerge as a key model of massively parallel
computation. Exact mathematical computations are possible, since one
can simulate a Turing machine, but cellular automata have a very
different way to represent data. The geometrical aspect of cellular
automata induces specific questions that do not appear in sequential
models.

Whereas the work of a CA is based on local exchange in the
nearest neighborhood, at global scale the collective behavior of the
CA often emerges as signals, i.e. continuous lines in the
space-time diagram, which capture the organization and the sending of
information through the network. Cellular automata as computational
systems can be seen from two main points of view: either a CA is
designed to fill a specific task, or a given CA is analyzed in terms
of general properties and dynamics. In both cases, the notion of
signal appears. To build a CA, signals are a tool that makes the
transition from the local to the global behavior, to geometrically
describe the organization and the motion of information between cells
(see e.g.~\cite{PF,JM}). When analyzing a CA, the behavior of many CA
shows ``particles in motions'', whose trajectories can be interpreted as
signals (see~\cite{BM}, or even the gliders in the game of
Life~\cite{LIFE}).

Intuitively, signals are some paths through the space-time diagram
which encode and combine the information, but an all-encompassing
formalization is lacking. Nevertheless, some attempt has been done
(see~\cite{JMVT}). We propose an alternative definition for CA that
generate signals.

In dimension $1$, it has been shown that some signals around the
diagonal axis can not be set up by any CA: a signal set up by any CA
either becomes parallel to the diagonal axis or takes at least a
logarithmic slow-down. Surprisingly, in higher dimensions, although
more cells are involved around the diagonal axis, we will show that
the same gap occurs. So, increasing the dimension does not help to
construct such signals around the diagonal axis. This partially
answers the problem \#51 of the list of open problems on CA
(see~\cite{DFM}).

However, we have a gain in terms of number of states. In dimension
$1$, performing along the diagonal axis a logarithmic slow-down
requires at least $4$ states (it is not difficult to review the few CA
with $3$ states). But in dimension $2$, we exhibit a CA with $3$
states (including the quiescent state) which performs a logarithmic
slow-down along the diagonal axis. Furthermore we show that this CA is
optimal in terms of number of states.

To complete the analysis of the gain of working in higher dimension,
we describe a CA that supports other logarithmic slow-downs with less
states in dimension $2$ than in dimension $1$.

\section{Definition of a signal}

A $k$-dimensional cellular automata is a $k$-dimensional array of
finite automata (cells) indexed by $\zed^k$. All cells evolve
synchronously at discrete time steps. At each step, each cell enters a
new state according to a transition function involving only its local
neighborhood.

We use the notation $\vectk{u}=\left(u_1,\dots,u_k\right)$ to
designate a $k$-vector. $\vectk{0}$ is the null vector
$\left(0,\dots,0\right)$. $\vectk{1}$ is the unary vector
$\left(1,\dots,1\right)$ and $t\cdot\vectk{u}$ is the product of
$\vectk{u}$ by a scalar $t$.

Formally a $k$-CA is defined by $\left(\states,V,f,\lambda\right)$
where: $\mathcal{S}$ is the set of states,
$V=\left\{\vectk{x}^1,\dots,\vectk{x}^v\right\}\subset\zed^k$ is the
neighborhood, $f$ from $\states^{v}$ into $\states$ is the transition
function, $\lambda\in\states$ is the quiescent state which verifies
$f\left(\lambda,\dots,\lambda\right)=\lambda$.

A site \cellk{\vectk{u}}{t} refers to the cell \vectk{u} at time $t$
and \statk{\vectk{u}}{t} denotes its state at time $t$. We refer to
the whole mapping $\cellk{\vectk{u}}{t}\mapsto\statk{\vectk{u}}{t}$ as
the space-time diagram of the CA.

For time $t\geq0$ we have
$$\statk{\vectk{u}}{t+1}=f\left(\statk{\vectk{u}+\vectk{x}^1}{t},\dots,%
  \statk{\vectk{u}+\vectk{x}^v}{t}\right)$$

We will consider three different neighborhoods: the Von Neumann
neighborhood, the Moore neighborhood and the trellis neighborhood.
$$V_{\text{Von Neumann}}=\left\{\vectk{x}\in\zed^k:\sum
  \left|x_{i}\right|\leq 1\right\},$$
$$\Vmoore=\left\{\vectk{x}\in\zed^k:\left|x_{i}\right|\leq 1\right\},$$
$$\Vtrellis=\left\{\vectk{x}\in\zed^k:\left|x_{i}\right|=1\right\}.$$

Note that, with the trellis neighborhood, the states
\statk{\vectk{u}}{t} and \statk{\vectk{u}'}{t'} do not interfere if for
some $i$ the sums $u_i+t$ and $u'_i+t'$ are not of same parity.  So at
time $t$ we will deal only with cells $\vectk{u}=(u_1,\cdots ,u_k)$
such that $u_1, \cdots,u_k, t$ are of same parity, the other sites
are considered as quiescent or non-existent.

Observe that the graph of dependencies of a $k$-dimensional cellular
automata with Moore neighborhood contains the graph of dependencies of
a $k$-dimensional cellular automata with Von Neumann neighborhood; so
the simulation of a Von Neumann CA can be done in real time by a Moore
CA. The graph of dependencies of a $k$-CA with Moore neighborhood also
contains the graph of dependencies of a $k$-dimensional trellis. And
as shown in dimension $1$ (see~\cite{CHCU,IKM}),
provided the cells $\vectk{u}$ of a CA with trellis neighborhood
correspond to the set of cells $\left\{\vectk{u}+\vectk{x}: \vectk{x}
  \in \{0,1\}^k\right\}$ of a CA with Moore neighborhood, the trellis
CA and the Moore CA are time-wise equivalent.  Hence a trellis CA
which performs the same task than a Moore CA, might have more states
but always with less interconnections.

We recall the definition of impulse CA's and signals (see~\cite{JMVT}):

\begin{Definition}[Impulse CA]
  An impulse CA is a $5$-tuple $\left(\states,V,f,G,\lambda\right)$
  where $\left(\states,V,f,\lambda\right)$ is a CA and $G$ a
  distinguished state of $\states$ such that at initial time $t=0$ all
  cells are in the quiescent state $\lambda$ but the cell $\vzero$
  which is in state $G$:
$$\left\{\begin{array}{l}%
\statk{\vectk{x}}{0}=\lambda\mbox{\quad if~}\vectk{x}\neq\vzero,\\%
\statk{\vzero}{0}=G.\end{array}\right.$$
\end{Definition}

\begin{Definition}[Signal]
  For a given neighborhood $V$, a \Vsignal $\Gamma$ is a sequence of
  sites $\left\{\cellk{\vectk{u}(t)}{t}\right\}_{t\geq 0}$ such that
\begin{itemize}
\item $\vectk{u}(0)=\vzero$.
\item For all $t\geq 0$: $\vectk{u}(t+1)-\vectk{u}(t)\in V$.
\end{itemize}
\end{Definition}
Fundamentally, a signal is a continuous path in the graph of
dependencies of the CA.

To emphasize the elementary moves of the \Vsignal $\Gamma$, we
denote by $\Gamma_\vectk{x}$ where $\vectk{x}\in V$, the set of sites
of $\Gamma$ which reach the next one by a $\vectk{-x}$ move:
$\Gamma_\vectk{x}=\left\{\cellk{\vectk{u}(t)}{t}\in\Gamma:
  \cellk{\vectk{u}(t)-\vectk{x}}{t+1}\in\Gamma\right\}$. Note that
$\left\{\Gamma_\vectk{x}\right\}_{\vectk{x}\in V}$ defines a partition
of $\Gamma$.

We recall the definition of impulse CA which draw explicitly a signal.

\begin{Definition}[Construction of a signal]
  An impulse CA $A=\left(\states,V,f,G,\lambda\right)$ constructs a
  \Vsignal $\Gamma$ if there
  exists a subset $\states_0$ of $\states$ such that
  $\cellk{\vectk{u}}{t}\in\Gamma$ if and only if
  $\statk{\vectk{u}}{t}\in\states_0$.
\end{Definition}

We propose also two alternative definitions of impulse CA which 
draw implicitly signals.

\begin{Definition}[Detection of a signal]
  An impulse CA $A=\left(\states,V,f,G,\lambda\right)$ detects a
  \Vsignal $\Gamma$ if there exists a partition
  $\left\{\states_\vectk{x}\right\}_{\vectk{x}\in V}$ of the set of
  states $\states$ such that if
  $\cellk{\vectk{u}}{t}\in\Gamma_\vectk{x}$ then
  $\statk{\vectk{u}}{t}\in\states_\vectk{x}$.
\end{Definition}

\begin{Definition}[Supporting a signal]
  An impulse CA $A=\left(\states,V,f,G,\lambda\right)$ supports a
  \Vsignal $\Gamma$ if there exists a finite automaton
  $F=\left(\states,Q,\delta,q_0\right)$ with $\states$ the input 
  alphabet, $Q$ the set of states, $\delta$ from $Q\times \states$ into
  $Q\times V$ the transition function and $q_0$ the initial state
  and a sequence of states $\left\{q(t)\right\}_{t\geq 0}$ such that
  $q(0)=q_0$ and
  for all $t\geq 0$: $\delta\left(q(t),\statk{\vectk{u(t)}}{t}\right)
  =\left(q(t+1),\vectk{u}(t+1)-\vectk{u}(t)\right)$.
\end{Definition}

The construction of a signal is a characterization by marking all the
sites of the signal with a special set of states, whereas supporting a
signal is a more dynamic tool, enabling the use of a finite automaton
to retrieve the signal from the space-time diagram. Detection is a
special case of support.

Actually the three notions are equivalent. If an impulse CA $A$
constructs a \Vsignal $\Gamma$ then it detects it and if an impulse CA
$A$ detects a \Vsignal $\Gamma$ then it supports it.  Furthermore, we
get:

\begin{Proposition}
  If an impulse CA $A$ supports a \Vsignal $\Gamma$ then
  there exists an impulse CA $A'$ which constructs it.
\end{Proposition}

\begin{proof}
  Suppose that $\Gamma$ is supported by the impulse CA
  $A=\left(\mathcal{S},V,f,G,\lambda\right)$ with the finite automata
  $F=\left(\states,Q,\delta,q_0\right)$.  Consider the new impulse CA
  $A'=\left(\states\times\left(\{0\}\cup Q\right),V,f',
    (G,q_0),(\lambda,0)\right)$ with
  $$f'(\underbrace{(s_\vectk{x},m_\vectk{x}),\dots}_{\vectk{x}\in
    V})=(f(\underbrace{s_\vectk{x},\dots}_{\vectk{x}\in V}),m))$$
  where $m\in Q$ if and only if there exists $\vectk{a}\in V$ such that
 $m_\vectk{a}\in Q$ and $\delta\left(m_\vectk{a},s_\vectk{a}\right)=%
 (m,\vectk{a})$. 
 Then the subset $\states_0=\states\times Q$ marks exactly the sites of
  $\Gamma$.
\end{proof}

\begin{Definition}[Basic signals]
  A \Vsignal is basic if the sequence of its elementary moves (whose
  values are in $V$)
  $\left\{\vectk{u}(t+1)-\vectk{u}(t)\right\}_{t\geq 0}$ is ultimately
  periodic.
\end{Definition}

Actually the basic signals do not use the parallelism of the CA:
\begin{Claim}\label{claim-basic-signals}
  The impulse CA
  $\mathcal{A}=\left(\{\lambda\},V,f,\lambda,\lambda\right)$ supports
  exactly the basic \Vsignal[s].
\end{Claim}

\begin{proof}
  Any impulse CA $\left(\states,V,f,\lambda\right)$, in particular the
  CA $\mathcal{A}=\left(\{\lambda\},V,f,\lambda,\lambda\right)$,
  supports any basic \Vsignal.  Conversely, a quiescent background can
  only support ultimately periodic moves.
\end{proof}
\section{A gap on constructible signals}

In dimension 1, it has been shown that the signals
$\left\{\cellk{t-u(t)}{t}\right\}_{t\geq 0}$ such that $u(t)=
o\left( \log (t) \right)$ and $u(t)\neq \Theta(1)$ 
are not constructible (see~\cite{JMVT}). Here we will show
for Moore neighborhood (and therefore trellis neighborhood)
that even in higher dimension, the
signal of maximal speed $\left\{\cellk{t\cdot\vun}{t}\right\}_{t\geq 0}$ 
can not be slowed down below the logarithm.

First we define, for $\vki\in\zed^k$ and $t\in\ene$, $\D{\vki}{t}$ to
be the state $\statk{t\cdot\vun-\vki}{t}$. The states of the neighbor
cells of $\D{\vki}{t+1}$ with relative coordinates $\xinVM$ are
$\D{\vixu}{t}$. Thus, $\D{\vki}{t+1}=\fff{t}$. And at initial time,
only the cell $\vzero$ is in a non-quiescent state $G$:
$$\D{\vki}0=\left\{\begin{array}{l}
    G\quad\text{if~}\vki=\vzero\\\lambda\quad\text{else.}\end{array}\right.$$
\begin{Claim}
  $\D{\vki}{t}=\lambda$ if $\vki\in\zed^k\setminus\ene^k$ or
  $2t<\max\left(i_{1},\dots,i_{k}\right)$.
\end{Claim}

\begin{proof}
As {\vzero} is the only active cell at time $0$, at time $t\geq 0$
(with Moore neighborhood) a cell $\vectk{c}$ is in a quiescent state
if any of its coordinates $c_a$ is such that $|c_a|>t$. In particular,
with $\vectk{c}=t\cdot\vun-\vki$, if any $i_a$ is such that
$|t-i_a|>t$, i.e. $i_a<0$ or $i_a>2t$, we have $\D{\vki}{t}=\lambda$.
\end{proof}
We consider the words $\Dg{\vki}\in\states^\infty$ corresponding to
the significant part of the diagonals:
$$\Dg{\vki}=\left\{\begin{array}{l}%
(\D{\vki}{t})_{t\geq\left\lceil{\max\left(i_{1},\dots,i_{k}%
\right)}{/2}\right\rceil}\quad\text{if~}\vki\in\ene^k\\%
\lambda^\infty\quad\text{else.}\end{array}\right.$$

The next proposition states the periodic behavior of \Dg{\vki}. As
$\D{\vki}{t+1}$ is defined by $\D{\vixu}{t}$ where $\xinVM$, the
periodic behavior of \Dg{\vki} can be characterized by the periodic
behavior of the lower diagonals \Dg{\vixu} where $\xinVMred$.

\begin{Proposition}\label{propperiod}
  For all $\vki\in\zed^k$, there exists
  $\alpha_{\vki}\in\states^\star$, $\beta_{\vki}\in\states^\star$,
  $u_{\vki}\in\ene$ and $v_{\vki}\in\ene$ such that:
  \begin{itemize}
  \item $\Dg{\vki}=\alpha_{\vki}\left(\beta_{\vki}\right)^\infty$.
  \item $u_{\vki}+v_{\vki}\leq \cardstates$ and $1 \leq v_{\vki}\leq 
    \cardstates$.
  \item $|\alpha_{\vki}|\leq M_{\vki}+u_{\vki} P_{\vki}$, where
    $M_{\vki}$ is
    $\displaystyle\max_{\xinVMred}\left(|\alpha_{\vixu}|\right)$.
  \item $|\beta_{\vki}|$ divides $v_{\vki}P_{\vki}$, where $P_{\vki}$
    is $\displaystyle\lcm_{\xinVMred}\left(|\beta_{\vixu}|\right)$.
\end{itemize}
\end{Proposition}

\begin{proof}
  We do an induction on $r=i_1+\dots+i_k$. Remark that the sum of all
  coordinates of $\vixu$ is always smaller than the sum of all
  coordinates of \vki, for $\xinVMred$. The proposition is true for
  $r<0$. Indeed in this case $\vki\in\zed^k\setminus\ene^k$. So
  $\Dg{\vki}$ is $\lambda^\infty$ and we can set $\alpha_{\vki}$ to be
  the empty word, $\beta_{\vki}=\lambda$, $u_{\vki}=0$ and
  $v_{\vki}=1$.  We suppose the proposition true up to $r$ and we will
  prove it for $r+1=i_1+\cdots+i_k$. Let $\tau$ stand for
  $\left\lceil{\max\left(i_{1},\dots,i_{k}\right)}/{2}\right\rceil$.
  By hypothesis of recurrence, we have $\D{\vixu}{t}=\D{\vixu}{t'}$
  for all $\xinVMred$ and all $t$, $t'$ such that $t>t'\geq \tau +
  M_{\vki}$ and $P_{\vki}$ divides $(t-t')$ ($P_{\vki}$ is the least
  common multiple of all the periods for $\xinVMred$, hence the
  periodicity).
  
  Among the $\cardstates+1$ states $\left\{\D{\vki}{\tau+M_{\vki}},
    \D{\vki}{\tau+M_{\vki}+P_{\vki}}, \dots,
    \D{\vki}{\tau+M_{\vki}+\cardstates P_{\vki}}\right\}$ at least two
  are equal: for some $a$ and $b$ with $0\leq a<b\leq\cardstates$,
  $\D{\vki}{\tau+M_{\vki}+aP_{\vki}}=\D{\vki}{\tau+M_{\vki}+bP_{\vki}}$.
  Thus, by induction, it follows that for all $u$:
  \begin{align*}
    \D{\vki}{\tau+M_{\vki}+aP_{\vki}+u+1}&=\fff{\tau+M_{\vki}+aP_{\vki}+u}
    =\fff{\tau+M_{\vki}+bP_{\vki}+u}\\&=\D{\vki}{\tau+M_{\vki}+bP_{\vki}+u+1}.
  \end{align*}
  
  We can choose
  $\alpha_{\vki}=\D{\vki}{\tau}\cdots\D{\vki}{\tau+M_{\vki}+aP_{\vki}-1}$,
  $\beta_{\vki}=\D{\vki}{\tau+M_{\vki}+aP_{\vki}}\cdots\D{\vki}%
  {\tau+M_{\vki}+bP_{\vki}-1}$,
  $u_{\vki}=a$ and $v_{\vki}=b-a$ which verify the desired properties.
\end{proof}

The following corollary specifies the length of the periodic and 
non-periodic parts of $\Dg{\vki}$.
\begin{Corollary}\label{corollaire-longueur}
  For all $\vki\in\ene^k$, there exists
  $\alpha_{\vki}\in\states^\star$, $\beta_{\vki}\in\states^\star$ such
  that
  \begin{itemize}
  \item $\Dg{\vki}=\alpha_{\vki}\left(\beta_{\vki}\right)^\infty$.
  \item $|\alpha_{\vki}|<\cardstates\lcms$.
  \item $|\beta_{\vki}|$ divides $\lcms[+1]$.
  \end{itemize}
\end{Corollary}

\begin{proof} We do a recurrence on $r=i_1+\cdots+i_k$. For $r=0$,
according the proposition we have $|\alpha_\vzero|\leq
u_\vzero<\cardstates$, $|\beta_\vzero|$ divides $ v_\vzero$ which
divides $\lcm(1,\dots,\cardstates)$.

Now we suppose the corollary true up to $r$. Then for
$r+1=i_1+\cdots+i_k$ we have $M_{\vki}<\cardstates\lcms[-1]\leq\lcms$;
$P_{\vki}$ divides $\lcms$. So
\begin{equation*}\begin{split}
    |\alpha_{\vki}|
    &\leq M_{\vki}+u_{\vki}P_{\vki}\\&<\lcms+(\cardstates-1)\lcms\\
    &<\cardstates\lcms.
\end{split}\end{equation*}
And $|\beta_{\vki}|$ divides $v_{\vki}P_{\vki}$ which divides $\lcms[+1]$.
\end{proof}

The following claim emphasizes a first constraint on constructible
signals implied by proposition~\ref{propperiod}.
\begin{Claim}\label{claim-constant}
  If a \Vsignal $\Gamma=\left\{\cellk{\vectk{u}(t)}{t}\right\}_{t\geq
    0}$, constructed by an impulse CA (with Moore neighborhood) 
  enters the periodic part of the
  CA at some step $t_0$ then the \Vsignal becomes constant: for all
  $t\geq t_0$, $\vectk{u}(t)=\vectk{u}(t_0)+(t-t_0)\cdot\vun$.
\end{Claim}

\begin{proof}
  The value of $\cellk{\vectk{u}(t_0)}{t_0}$ belongs to the subset
  $\states_0$ which marks the \Vsignal $\Gamma$. Moreover, it belongs
  to the diagonal $\Dg{\vs}$. Suppose this site belongs to the
  periodic part $\beta_{\vs}$. Then from $\cellk{\vectk{u}(t_0)}{t_0}$
  onward, there is an infinite number of sites of $\Dg{\vs}$ which
  belong to $\states_0$. As $\Gamma$ must go through all sites whose
  states belong to $\states_0$, the signal $\Gamma$ always remains on
  the diagonal $\Dg{\vs}$.
\end{proof}

Finally we exhibit the gap relating to constructible signals.
\begin{Proposition}\label{prop-gap}
Let a \Vsignal $\left\{\cellk{t-\vectk{u}(t)}{t}\right\}_{t\geq 0}$ and
\\$m(t)=\max\left(u_{1}(t),\dots,u_{k}(t)\right)$
be such that:
\begin{itemize}
\item $m(t)$ is not constant: $m(t)\neq \Theta (1)$;
\item $m(t)$ is below the logarithm: $m(t)= o(\log(t))$.
\end{itemize}
Then there exists no impulse CA with Moore neighborhood
which supports such \Vsignal.
\end{Proposition}

\begin{proof}
  According to claim~\ref{claim-constant}, a \Vsignal
  $\left\{\cellk{t-\vectk{u}(t)}{t}\right\}_{t\geq0}$ constructible by
  an impulse CA with Moore neighborhood , providing $m(t)\neq \Theta
  (1)$, never enters the periodic part of the CA.  Moreover, observe
  that $\statk{t-\vectk{u}(t)}{t}$ belongs to $\Dg{\vectk{u}(t)}$.
  
  The non-quiescent part of $\Dg{\vectk{u}(t)}$ begins on the site
  $\cellk{\left\lceil{m(t)}/{2}\right\rceil\cdot\vun
    -\vectk{u}(t)}{\left\lceil{m(t)}/{2}\right\rceil}$ and so the
  periodic part of $\Dg{\vectk{u}(t)}$ begins on the site
  $\cellk{\left(\left\lceil{m(t)}/{2}\right\rceil+%
      |\alpha_{\vectk{u}(t)}|\right)\cdot\vun-\vectk{u}(t)}%
  {\left\lceil{m(t)}/{2}\right\rceil+|\alpha_{\vectk{u}(t)}|}.$ Hence
  if the signal is constructible, we get for all $t$, $t <
  \left\lceil{m(t)}/{2}\right\rceil+|\alpha_{\vectk{u}(t)}|$; and
  according to corollary~\ref{corollaire-longueur}, $t <
  \left\lceil{m(t)}/{2}\right\rceil+\cardstates^{1+k\cdot m(t)}$.  So
  for some constant $C$, we have for all $t$: $t < C^{m(t)}$.  In
  other words $m(t)= \Omega\left( \log (t) \right)$.
\end{proof}
\noindent\textbf{Remark 1.}
Remark that the periodic phenomenon we just have examined along the
signal of maximal speed $\left\{\cellk{t\cdot\vun}{t}\right\}_{t\geq
  0}$, by an adequate rotation, occurs along all signals
$\left\{\cellk{t\cdot\vectk{x}}{t}\right\}_{t\geq 0}$ with
$\vectk{x}\in\Vtrellis$. The proposition~\ref{prop-gap} remains true
for any \Vsignal $\left\{\cellk{\vectk{c}(t)}{t}\right\}_{t\geq0}$ and
$m(t)=\max\left(|c_{1}(t)|-t,\dots,|c_{k}(t)|-t\right)$ with $m(t)\neq
\Theta (1)$ and $m(t)= o(\log(t))$.

\noindent\textbf{Remark 2.} Due to the equivalence of Moore CA and trellis CA,
the same limitation operates for the construction of $\Vtrellis$-signals
on CA with trellis neighborhood.

\noindent\textbf{Remark 3.} With von Neumann neighborhood, the signals 
$\left\{\cellk{\left(t-m(t)\right)\cdot\vectk{x}}{t}\right\}_{t\geq0}$
with $\vectk{x}\in V_{\text{Von Neumann}}$, $m(t)\neq \Theta (1)$ and
$m(t)=o(\log(t))$ are not constructible by any impulse CA with Von
Neumann neighborhood; otherwise using an adequate rotation, signals
such $\left\{\cellk{\left(t-m(t)\right)\cdot\vun}{t}\right\}_{t\geq0}$
would be constructible by an impulse CA with Moore neighborhood,
contradicting proposition~\ref{prop-gap}.

\section{Construction of the logarithm in dimension 2}
Let $\mathcal{A}$ be the following impulse $2$-CA with the
neighborhood
$\Vtrellis=\left(\xx{1}{1},\xx{-1}{-1},\xx{1}{-1},\xx{-1}{1}\right)$.
The set of states $\mathcal{S}$ is $\{\lambda,0,1\}$, the initial
distinguished state is $1$, the quiescent state is $\lambda$ and the
transition function $f$ is depicted in
figure~\ref{fonction-deuxetats}.
\begin{figure}
$$\begingroup\newcounter{tabline}\setcounter{tabline}{-1}%
\def\increasetabline{\stepcounter{tabline}\mbox{\#\thetabline}}
\catcode`\ =4\catcode`\*=\active\let*=\star%
\catcode`\!=\active\let!=\increasetabline%
\catcode`\?=\active\let?=\lambda%
\begin{array}{|cccc|c|c|}\hline%
a b c d f(a,b,c,d) \mbox{Rule\space{}number}\\\hline%
? ? ? ? ? !\\
1 ? ? ? 0 !\\
0 ? ? ? 1 !\\
? ? 0 1 1 !\\
1 ? 0 1 0 !\\
0 ? 0 1 1 !\\
1 ? 1 0 1 !\\
1 ? 0 0 1 !\\
0 ? 1 0 0 !\\
0 ? 0 0 0 !\\
* 1 ? * 1 !\\
* 1 1 * 1 !\\
* 1 0 * 0 !\\
* 0 * * 0 !\\
* * * * ? !\\
\hline\end{array}\endgroup\quad\parbox{6cm}{%
$a$, $b$, $c$ and $d$ are the cells with the following relative
coordinates in the space-time diagram:\begin{itemize}\item $a$ is
  \xxx{-1}{-1}{-1},\item $b$ is \xxx{-1}{1}{-1},\item $c$ is
  \xxx{1}{1}{-1}\item $d$ is \xxx{1}{-1}{-1}.\end{itemize}Rules are
sorted by order of precedence.}$$
\caption{Transition function for $\mathcal{A}$.}
\label{fonction-deuxetats}
\end{figure}
\newcommand{\lo}[1]{\log_2(#1+1)}

\begin{Proposition}\label{proposition-deux-etats}
  Let $\ell$ be the function
  $t\mapsto\left\lfloor\log_2(t+1)\right\rfloor$. $\mathcal{A}$
  detects the signal
  $\left\{\xxx{t-\ell(t)}{t-\ell(t)}{t+\ell(t)}\right\}_{t\geq 0}$,
  with the following partition: $\states_{\xx{1}{1}}=\{0\}$ and
  $\states_{\xx{-1}{-1}}=\{1\}$.
\end{Proposition}

\begin{proof}
\begin{Claim}
  All cells with state $1$ or $0$ (called ``active cells'') have
  coordinates \xxx xyz such that $-z\leq y\leq x\leq z$, $z\geq 0$ and
  $x+z$ and $y+z$ are both even (``even'' cells).
\end{Claim}

The property of $x+z$ and $y+z$ is always true in such a trellis (see
the definition). The condition $z\geq 0$ is quite obvious, as this is
mandated by the definition of a space-time diagram. Now, let us look
at the relation $-z\leq x\leq z$ ($-z\leq y\leq z$ can be proved by
the same arguments). As the only cell at time $z=0$ that has a
non-quiescent state has the property that $x=0$, and that the
neighborhood's $x$-range is $\{1, -1\}$, no cell can enter a
non-quiescent state unless it includes the cell \xxx 000 in its
dependencies, i.e. if $-z\leq x\leq z$.
  
The remaining condition $y\leq x$ is due to the transition function.
Let us suppose that the condition is true up to some value $z$, and
let us check that the condition at $z+1$ still holds true, i.e. that
no rule numbered from \#1 to \#13 is applied to any cell such that
$x>y$. If the rule holds true up to $z$, then all neighbors except the
one with relative coordinates \xxx 1{-1}{-1} are such that $x>y$. So,
their states are the quiescent state $\lambda$. Thus, the only
possibility for cells such that $x>y$ is to meet the quadruplet
$(\lambda,\lambda,\lambda,1)$ or $(\lambda,\lambda,\lambda,2)$ or the
quiescent rule (\#0). As both these quadruplets fall in the catch-all
rule (\#14), the induction is proved (it is obviously true for $z=0$).
  
Let us now define $\overline{k}$ by $k=\sum_i \overline{k}^{(i)} 2^i$)
(i.e. $\overline{k}$ is the binary writing of $k$). We also define
$\tilde{k}$ to be the number of consecutive $1$'s at the beginning of
$\overline{k}$. Let us define a spatial transformation for the
space-time diagram with:
  
$$\W{k}{l}{i}=\valeur{k-i+l}{k-i-l}{k+i+l}.$$
  
Let us call \W[A]{k}{l}{i}, \W[B]{k}{l}{i}, \W[C]{k}{l}{i} and
\W[D]{k}{l}{i} the four neighbors of cell \W{k}{l}{i}. We have the
following equalities:
$$\W[A]{k}{l}{i}=\W{k-1}{l}{i}\quad \W[B]{k}{l}{i}=\W{k}{l-1}{i}\quad
\quad\W[C]{k}{l}{i}=\W{k}{l}{i-1}
\quad\W[D]{k}{l}{i}=\W{k-1}{l+1}{i-1}$$
  
As proven by claim 1, \W{k}{l}{i} is $\lambda$ if any of the indices
is $-1$. We consider the words \WW{k}{l} that we get by the
concatenation of all \W{k}{l}{i}, for $i\geq 0$:

  \begin{Claim}
    \WW{k}{0} is $\overline{k+1}\lambda^\infty$, and for all $l>0$,
    \WW{k}{l} is
    $1^{\widetilde{k+1}}0^{|\overline{k+1}|-\widetilde{k+1}}\lambda^\infty$.
  \end{Claim}
  
  The execution can be seen on figure~\ref{figure-graphique}. In fact,
  we can read the binary writing of $k+1$ according to some spatial
  transformation (the one performed by $\W kli$).
\begin{figure}
  \noindent\centerline%
  {\includegraphics[angle=90,height=0.6\textwidth,origin=c]{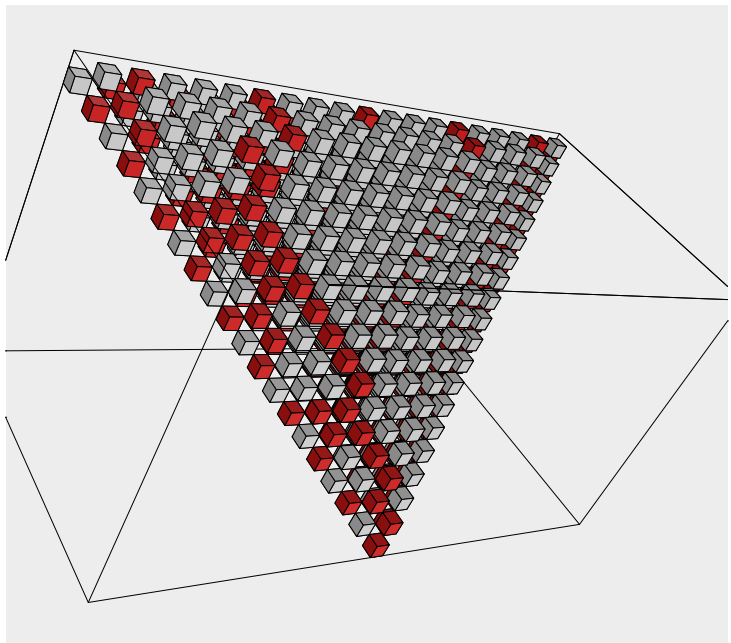}}
\caption{Sample execution of $\mathcal{A}$. White cubes are sites in%
  state $0$, dark cubes are sites in state $1$. The binary writing%
  can be seen horizontally on the picture. Depth is difficult to perceive.}
\label{figure-graphique}
\end{figure}

We shall prove the claim with an induction on $k$, and dividing the
proof in two subcases: whether $k=2^j-2$ for some $j$ or not.
  
  \textbf{Subcase 1:} $k=2^j-2$. We suppose the claim true up to $k$,
  and we shall prove the claim for $k+1$. The claim can be transformed
  this way for $k=2^j-2$: $\W{k}{l}{i}=1$ for any $l$ (even $l=0$) and
  $0\leq i<j$, else $\W{k}{l}{i}=\lambda$. We get the following items:
  \begin{itemize}
  \item $\W[B]{k+1}{0}{i}=\lambda$ for all values of $i$ (see claim
    1).
  \item $\W{k+1}{0}{0}=0$ (according to rule \#1).
  \item With a quick induction (for \W[C]{k+1}{0}{m}),
    $\W{k+1}{0}{m}=f(1,\lambda,0,1)=0$ for all $m<j$ (rule \#4).
  \item
    $\W{k+1}{0}{j}=f(\W{k}{0}{j},\lambda,\W{k+1}{0}{j-1},%
    \W{k}{1}{j-1})=f(\lambda,\lambda,0,1)=1$
    (rule \#3).
  \item
    $\W{k+1}{0}{j+1}=f(\W{k}{0}{j+1},\lambda,\W{k+1}{0}{j},%
    \W{k}{1}{j})=f(\lambda,\lambda,1,\lambda)=\lambda$
    (rule \#14).
  \item With a quick induction (for \W[C]{k+1}{0}{m}),
    $\W{k+1}{0}{m}=\lambda$ for all $m>j+1$ (rule \#0).
  \item Now, we use an induction on $l$, since we proved the claim for
    $l=0$. $\W[B]{k+1}{l}{i}=0$ for all values of $i$ such that $0\leq
    i\leq j$, except if $l=1$ (where $\W[B]{k+1}{1}{j}=1$). So, rule
    \#13 applies for all $i$ such that $0\leq i\leq j$ and
    $\W{k+1}{l}{i}=0$. The only exception is
    $\W{k+1}{1}{j}=f(\lambda,1,0,1)=0$ according to rule \#12.
  \item We continue the induction on $l$ to use rule \#14 for
    $\W{k+1}{l}{j+1}$.
    $$\W{k+1}{l}{j+1}=f(\W{k}{l}{j+1},\W{k+1}{l-1}{j+1},\W{k+1}{l}{j},
    \W{k}{l+1}{j})=f(\lambda,\lambda,0,\lambda)=\lambda.$$
  \item Last, we compute the values for $l>0$ and $i>j+1$. All
    neighbors are $\lambda$, therefore $\W{k+1}{l}{i}=\lambda$.
  \end{itemize}
  
  There is one special case, the case $k=0$, which is the starting
  point of the induction. $\W 000$ is obviously what we are looking
  for. For all values of $l>0$, we have
  $\W{0}{l}{0}=f(\lambda,\W{0}{l-1}{0},\lambda,\lambda)=1$, by using
  rule \#10. For all other $\W{0}{l}{i}$, either rule \#14 or \#0 will
  be used, the result being always $\lambda$.

  \textbf{Subcase 2:} $\forall j,k\neq 2^j-2$. As such, there is at
  least one $0$ in the binary writing of $k+1$. Let us recapitulate a few
  facts about the incrementation of a binary number: to increment a
  binary number, all the initial (starting from lower-weight digits)
  $1$'s have to be turned into $0$'s, and the first $0$ is turned into
  a $1$ (the special case of $2^j-1$ will not appear there). Our first
  goal will be to check that $\WW{k+1}{0}$ is correctly generated from
  $\WW{k}{0}$. Remember that $\tilde{k}$ is the number of initial
  $1$'s in the binary writing of $k$. We prove the claim by induction:
  \begin{itemize}
  \item $\W[B]{k+1}{0}{i}=\lambda$ for all values of $i$ (see claim
    1). If $k$ is even then $\W{k+1}{0}{0}=0$ (according to rule \#1).
    If $k$ is odd, then $\W{k+1}{0}{0}=1$ (rule \#2).
  \item For $0<i\leq\widetilde{k+1}$, the value of
    $\W[D]{k+1}{0}{i}=1$. With a quick induction, we can prove that
    $\W{k+1}{0}{i}=0$ (rule \#4) as all $\W[A]{k+1}{0}{i}=1$. This
    condition is not met if $k$ is odd.
  \item For $i=\widetilde{k+1}$, rule \#5 is triggered
    ($\W{k+1}{0}{i}=1$). This is not done if $k$ was odd (in fact the
    incrementation is already over if $k$ is odd).
  \item For $\widetilde{k+1}<i<|\overline{k+1}|$ (and there is at
    least one value of $i$ for which this is true), the value of
    $\W[D]{k+1}{0}{i}=0$. As $\W[B]{k+1}{0}{i}=\lambda$, any of the
    rules \#6, \#7, \#8 or \#9 will be used. All those rules state
    that $\W{k+1}{0}{i}=\W[A]{k+1}{0}{i}=\W{k}{0}{i}$.
  \item For $i=|\overline{k+1}|$, we have
    $\W[A]{k+1}{0}{i}=\W[B]{k+1}{0}{i}=\lambda$, $\W[C]{k+1}{0}{i}=1$
    (since all binary writings end with a $1$). So, rule \#14 applies,
    and $\W{k+1}{0}{i}=\lambda$.
  \item For $i>|\overline{k+1}|$, a quick induction shows that
    $\W[C]{k+1}{0}{i}$ is $\lambda$, thus making $\W{k+1}{0}{i}$ be
    $\lambda$.
  \item We proved the claim for $l=0$ (see the preliminary explanation
    on binary incrementation). As $\W[C]{k+1}{l}{0}=\lambda$, it's
    easy to prove that for all $l>0$, $\W[B]{k+1}{l}{0}$ is not
    $\lambda$, so $\W{k+1}{l}{0}=\W{k+1}{0}{0}$ (using either rule
    \#10 or rule \#13).
  \item Let us consider now the case $0<i<\widetilde{k+2}$ (this case
    may not happen if $k$ is even). We prove by induction on $X=l+i$
    that in this case, the value is always $1$.
    $\W{k+1}{l}{i}=f(\W[A]{k+1}{l}{i},1,1,\W[D]{k+2}{l}{i})$, since
    $l-1$ and $i-1$ will both have a smaller sum than $X$. Thus, this
    sub-proof is done (using rule \#11, the value is always $1$). The
    proof for the case $l=1$ or $i=1$ is very easy (using the previous
    item).
  \item Now, let us consider the case where $i=\widetilde{k+2}$, with
    $i\neq 0$ and $i\neq|\overline{k+2}|$ (i.e. $k$ is odd and there
    is at least a $0$ in the binary writing of $k+2$). With a quick
    induction on $l$, as $\W{k+1}{0}{i}$ is $0$, we have
    $\W{k+1}{l}{i}=0$ for all $l$ (using rule \#13) (this is the first
    $0$ in the binary writing of $k+2$).
  \item Now, we study the case $\widetilde{k+2}<i<|\overline{k+2}|$.
    We still do an induction of $l+i$ and prove that the value is
    always $0$.
    $\W{k+1}{l}{i}=f(\W[A]{k+1}{l}{i},\W[B]{k+1}{l}{i},0,\W[D]{k+2}{l}{i})$,
    with \W[B]{k+1}{l}{i} being $0$ or $1$ (if $l=1$). Either way,
    rule \#12 or \#13 is used, and the value still ends up being $0$.
  \item We consider $i=|\overline{k+2}|$, and increasing values of
    $l$. $\W[A]{k+1}{l}{i}=\lambda$. \W[B]{k+1}{1}{i} is also
    $\lambda$. We will prove with an induction on $l$ that
    $\W[B]{k+1}{l}{i}=\lambda$ for any $l>0$. Let us presume it's
    true. $\W[C]{k+1}{l}{i}$ may be $0$, but
    $\W[D]{k+1}{l}{i}=\W{k}{l+1}{i-1}$ is always $0$. If it was not
    $0$, then $k+1$ would have no $0$ in its writing, and this is
    excluded in this subcase. So, rule \#3 does not apply, hence the
    result (rule \#14 is used).
  \item For larger values of $i$, the claim is straightforward,
    since only rule \#0 will be used.
\end{itemize}
\end{proof}

\begin{Proposition}[Optimality]
  The result of proposition~\ref{proposition-deux-etats} is optimal
  for dimension $2$, that means it is not possible to detect the
  signal $\left\{\xxx{t-\ell(t)}{t-\ell(t)}{t+\ell(t)}\right\}_{t\geq
    0}$ with only two states.
\end{Proposition}
\begin{proof}
  Let us suppose that there exists an impulse CA with only two states
  $0$ and $1$ contradicting the proposition.  Referring to
  claim~\ref{claim-basic-signals}, the general state of the impulse CA
  must not be the quiescent state. Let us decide that $0$ is the
  quiescent state. As we want to detect the aforementioned signal,
  then we can only choose one partition (because the state $1$ has to
  be in $\Gamma_{\xx{-1}{-1}}$). Thus, we have
  $\Gamma_{\xx{1}{1}}=\{0\}$ and $\Gamma_{\xx{-1}{-1}}=\{1\}$.
  
  Let us now consider a few sites of the signal. We must obtain:
  $$\valeur 000=1\quad\valeur 111=0\quad\valeur 002=1\quad\valeur
  113=1.$$
  
  Recall that $\valeur xyz$ is $0$ if $|x|>z$ or $|y|>z$.  We can
  rewrite two of these values the way they are computed from $f$:
  \begin{align*}
    \valeur 111&=f(\valeur 000,\valeur 200,\valeur 220,\valeur 020)=%
    f(1,0,0,0)=0\\
    \intertext{Thus, we get $f(1,0,0,0)=f(0,0,0,0)=0$. Now we write:}
    \valeur 113&=f(\valeur 002,\valeur 202,\valeur 222,\valeur 022)=%
    f(1,a,b,c)=1\\
    \intertext{with the following values for $a$, $b$ and $c$:}
    a=\valeur 202&=f(\valeur 1{-1}1,\valeur 3{-1}1,\valeur 311, \valeur 111)=%
    f(\valeur 1{-1}1,0,0,0)=0\\
    b=\valeur 222&=f(\valeur 111,\valeur 311,\valeur 331,\valeur 131)=%
    f(0,0,0,0)=0\\
    c=\valeur 022&=f(\valeur {-1}11,\valeur {-1}31,\valeur 131,
    \valeur 111)=f(\valeur {-1}11,0,0,0)=0
  \end{align*}
  
  Thus, $a=b=c=0$, and $\valeur 113=f(1,0,0,0)=0$. This is in
  contradiction with the fact that $\valeur 113$ must be $1$.
\end{proof}

\section{Building non-primal logarithmic signals}
Recall that, in dimension $1$, to build a logarithmic slow-down (in
base $b$) requires at least $b$ states. If $b$ is not primal, then it
can be written as the product of two numbers $x$ and $y$ whose gcd is
1. Here, we exhibit a $2$-dimensional CA that supports such a
logarithmic slow-down with only $x+y+2$ states instead of $xy$.

Let $x$ and $y$ be two integers such that $\gcd(x,y)=1$. Let
$\mathcal{A}$ be the following impulse $2$-CA with the neighborhood
$\Vtrellis=\left(\xx{1}{1},\xx{-1}{-1},\xx{1}{-1},\xx{-1}{1}\right)$.
The set of states $\mathcal{S}$ is
$\{\lambda,\pi_0,\pi_1,\dots,\pi_x,\kappa_0,\kappa_1,\dots,\kappa_y\}$,
the initial distinguished state is $\pi_1$, the quiescent state is
$\lambda$ and the transition function $f$ is as in
figure~\ref{fonction-xy}.
\begin{figure}
$$\begingroup\setcounter{tabline}{-1}%
\def\increasetabline{\stepcounter{tabline}\mbox{\#\thetabline}}
\catcode`\*=\active\let*=\star%
\catcode`\!=\active\let!=\increasetabline%
\catcode`\?=\active\let?=\lambda%
\begin{array}{|lccc|ll|c|}\hline%
A&B&C&D&\multicolumn{2}{|l|}{f(A,B,C,D)}&\mbox{Rule\space{}number}\\\hline%
?&?&?&?&?&&!\\%
\hline\multicolumn{7}{|c|}{\mbox{Rules for $l=0$}}\\\hline%
\pi_j&?&?    &?           &\pi_{j+1}&\text{(or $\pi_1$ if $j=k$)}&!\\%
\pi_x&?&\pi_k&\kappa_*    &\pi_0    &(k\neq x)&!\\%
\pi_j&?&\pi_k&\kappa_*    &\pi_j    &(j,k\neq x)&!\\%
\pi_x&?&\pi_x&\kappa_k    &\pi_0    &(k\neq y-1)&!\\%
\pi_j&?&\pi_x&\kappa_k    &\pi_j    &(j\neq x,k\neq y-1)&!\\%
\pi_j&?&\pi_x&\kappa_{y-1}&\pi_{j+1}&\text{(or $\pi_1$ if $j=k$)}&!\\%
?    &?&\pi_x&\kappa_{y-1}&\pi_1    &&!\\%
\hline\multicolumn{7}{|c|}{\mbox{Rules for $l=1$}}\\\hline%
\kappa_{y-1}&\pi_x&?,\kappa_y&?&\kappa_y    &         &!\\%
\kappa_{y-1}&\pi_k&?,\kappa_y&?&\kappa_0    &(k\neq x)&!\\%
\kappa_y    &\pi_*&?,\kappa_y&?&\kappa_{1}  &         &!\\%
\kappa_j    &\pi_*&?,\kappa_y&?&\kappa_{j+1}&(j\neq y-1,y)&!\\%
\kappa_y    &\pi_*&\kappa_k  &?&\kappa_0    &(k\neq y)&!\\%
\kappa_j    &\pi_*&\kappa_k  &?&\kappa_j    &(j\neq y,k\neq y)&!\\%
?           &\pi_1&?,\kappa_y&?&\kappa_{1}  &&!\\%
\hline%
*&*&*&*&?&&!\\
\hline\end{array}\endgroup\quad\parbox{35mm}{\sloppy%
$a$, $b$, $c$ and $d$ are the cells with the following relative
coordinates in the space-time diagram:\begin{itemize}\item $a$ is
  \xxx{-1}{-1}{-1},\item $b$ is \xxx{-1}{1}{-1},\item $c$ is
  \xxx{1}{1}{-1}\item $d$ is \xxx{1}{-1}{-1}.\end{itemize}Please note
that $\pi_\star$ is any state $\pi_j$, $\kappa_\star$ is any state
$\kappa_j$, and $\star$ is any state. Rules are sorted by order of
precedence.}$$
\caption{Transition function for $\mathcal{A}$.}
\label{fonction-xy}
\end{figure}

\begin{Proposition} Let $\ell$ be the function
  $t\mapsto\left\lfloor\log_{xy}(t+1)\right\rfloor$. Then
  $\mathcal{A}$ supports the signal
  $\left\{\xxx{t-\ell(t)}{t-\ell(t)}{t+\ell(t)}\right\}_{t\geq 0}$,
  using the following finite automaton:
\begin{gather*}
F=(\mathcal{S},\{a_1,\dots,a_y\}, \delta, a_1)\quad\text{with}\\
\left\{\begin{array}{l}\delta(a_y,\pi_x)=(a_1,\xx{1}{1})\\
\forall j\neq y,\delta(a_j,\pi_x)=(a_{j+1},\xx{-1}{-1})\\
\forall i\neq x,\forall j,\delta(a_j,\pi_i)=(a_j,\xx{-1}{-1})
\end{array}\right.
\end{gather*}
\end{Proposition}

\begin{proof}
Let us use the same spatial transformation as in the preceding section.
$$\W{k}{l}{i}=\valeur{k-i+l}{k-i-l}{k+i+l}.$$
Let us call \W[A]{k}{l}{i}, \W[B]{k}{l}{i}, \W[C]{k}{l}{i} and
\W[D]{k}{l}{i} the four neighbors of cell \W{k}{l}{i}. We have the
following equalities:
$$\W[A]{k}{l}{i}=\W{k-1}{l}{i}\quad
\W[B]{k}{l}{i}=\W{k}{l-1}{i}\quad
\quad\W[C]{k}{l}{i}=\W{k}{l}{i-1}
\quad\W[D]{k}{l}{i}=\W{k-1}{l+1}{i-1}$$

We have the following fact: only cells with $l=1$ or $l=0$ will be
non-quiescent. In fact, cells with $l=0$ will only have states in
$\pi_0,\dots,\pi_x,\lambda$ and cells with $l=1$ will only have states
in $\kappa_0,\dots,\kappa_y,\lambda$. This is quite easy to check:
rules \#1 to \#7 always require the neighbor $b$ to be $\lambda$, and
rules \#8 to \#14 always require $d$ to be $\lambda$.

The proof is on the same lines as the preceding proof: we can read the
writing of $k+1$ with the values of $\WW{k}{l}$. However, the writing is
a bit more complicated. In fact, $x$ and $y$ define a mapping of
$\{0,\dots,x-1\}\times\{0,\dots,y-1\}$ into $\{0,\dots,xy-1\}$ through
the Chinese remainder lemma. Let us call this mapping $\mu$, with the
added fact that $\pi_0$ and $\pi_x$ are made equivalent (idem for
$\kappa_0$ and $\kappa_y$). Then, we can read the writing in base $xy$
of $k+1$ by considering the $i$-th bit to be
$\mu(\W{k}{0}{i},\W{k}{1}{i})$.

The proof is quite cumbersome, and is reminiscent of the proof of
proposition~\ref{proposition-deux-etats}. The main point is that when
$\W k1i=\kappa_y$ is used instead of $\kappa_0$, it conveys the
information that the neighbor $\W k1i$ is exactly $\pi_x$, and thus
that the digit $\mu(\W k0{i+1},\W k1{i+1})$ has to be increased. The
fact that $\W k0i$ is $\pi_x$ instead of being $\pi_0$ carries the
fact that the corresponding $\W k1i$ has to be increased by $1$.
\end{proof}

\textbf{Possible enhancement:} It is possible to use the same set of
states for $\pi$ and $\kappa$. That is, the CA has just a set of
states $\{\pi_0,\dots,\pi_{\max(x,y)},\lambda\}$. It is necessary that
$x$ is the smallest of the two numbers $x$ and $y$. The transformation
of the transition function is as follows: each $\kappa_i$ becomes
$\pi_i$, and each $\lambda$ in the column $D$ of
figure~\ref{fonction-xy} becomes a $\star$ (any state).

\section{Prospectives}
Note that limitations in the construction of signals are likely
correlated to limitations in terms of language recognition. In
particular, the hierarchy between time $n$ and $n+\log n$ set up for
one-way cellular automata in dimension $1$ (see~\cite{KK}) might be
generalized to higher dimensions.

It should be possible to extend the last proposition in dimension $k$
as follows: one can get a $\log_{a}$ slow-down with the sum of $k$
factors whose $\gcd$ is 1 and whose product is $a$, plus $2$ (the
distinguished $\pi_x$ state and $\lambda$). Thus, the number of states
needed to get a $\log_a$ slow-down in dimension $k$ looks strongly
related to the decomposition of $a$ in prime numbers and the number of
its factors.


\begin{thebibliography}{DFM00}

\bibitem[BCG82]{LIFE}
E.~R. Berlekamp, John~H. Conway, and R.~K. Guy.
\newblock {\em Winning Ways for Your Mathematical Plays}, volume~2, chapter~25.
\newblock Academic Press, 1982.

\bibitem[C{\v C}84]{CHCU}
Christian Choffrut and Karel {\v C}ulik, II.
\newblock On real-time cellular automata and trellis automata.
\newblock {\em Acta Informatica}, 21:393--407, 1984.

\bibitem[DFM00]{DFM}
Marianne Delorme, Enrico Formenti, and Jacques Mazoyer.
\newblock Open problems on cellular automata.
\newblock Research report 2000-25, {\'E}cole normale sup{\'e}rieure de Lyon,
  July 2000.

\bibitem[Fis65]{PF}
Patrick~C. Fischer.
\newblock Generation of primes by a one-dimensional real-time iterative array.
\newblock {\em JACM}, 12:388--394, 1965.

\bibitem[IKM85]{IKM}
Oscar~H. Ibarra, Sam~M. Kim, and Shlomo Moran.
\newblock Sequential machine characterizations of trellis and cellular automata
  and applications.
\newblock {\em SIAM J. Comput.}, 14(2):426--447, 1985.

\bibitem[KK01]{KK}
Andreas Klein and Martin Kutrib.
\newblock A time hierarchy for bounded one-way cellular automata.
\newblock In J.~Sgall, A.~Pultr, and P.~Kolman, editors, {\em MFCS 2001},
  volume 2136 of {\em LNCS}. Springer, 2001.
\newblock To appear.

\bibitem[Mar00]{BM}
Bruno Martin.
\newblock Apparent entropy of cellular automata.
\newblock {\em Complex Systems}, 12(2), 2000.

\bibitem[Maz87]{JM}
Jacques Mazoyer.
\newblock A six-states minmal solution to the firing squad synchronization
  problem.
\newblock {\em TCS}, 50:183--238, 1987.

\bibitem[MT99]{JMVT}
Jacques Mazoyer and V{\'e}ronique Terrier.
\newblock Signals in one-dimensional cellular automata.
\newblock {\em TCS}, 217:53--80, 1999.

\bibitem[vN66]{vN}
John von Neumann.
\newblock {\em Theory of Self-Reproducing Automata}.
\newblock University of Illinois, Urbana, 1966.

\end{thebibliography}
\end{document}